\documentclass[aps,prd,superscriptaddress]{revtex4}
\usepackage{epsfig,epsf}
\usepackage{amsmath}
\usepackage{amsthm}
\usepackage{amsfonts}
\usepackage{amssymb}
\usepackage{dsfont}
\usepackage{multirow}
\usepackage{appendix}
\usepackage{slashed}
\usepackage[active]{srcltx}
\usepackage{psfrag}

\begin{document}

\title{Strong Decay of $P_c(4380)$  Pentaquark  in a Molecular Picture  }
\date{\today}
\author{K.~Azizi}

\affiliation{Physics Department, Do\u gu\c s University,
Ac{\i}badem-Kad{\i}k\"oy, 34722 Istanbul, Turkey}
\affiliation{School of Physics, Institute for Research in Fundamental Sciences (IPM), P. O. Box 19395-5531, Tehran, Iran}
\author{Y.~Sarac}
\affiliation{Electrical and Electronics Engineering Department,
Atilim University, 06836 Ankara, Turkey}
\author{H.~Sundu}
\affiliation{Department of Physics, Kocaeli University, 41380 Izmit, Turkey}

\begin{abstract}
 There are different assumptions on the substructure of the pentaquarks $P_c(4380)$ and $P_c(4450)$, newly founded in $J/\psi N  $ invariant mass by the LHCb collaboration, giving  consistent mass results with the experimental observations. The experimental data and recent theoretical studies on their mass suggest interpretation of  these states as spin-3/2 negative-parity and spin-5/2 positive-parity pentaquarks, respectively. There may exist  opposite-parity states corresponding to these particles, as well.  Despite a lot of studies, however, the nature and internal organization of these pentaquarks in terms of quarks and gluons are not clear.  To this end we need more theoretical investigations on other physical properties of these states. In this accordance, we study a strong and dominant decay of the $P_c(4380)$ to $J/\psi$ and $N$ in the framework of three point QCD sum rule method. An interpolating current in a molecular form is applied to calculate six strong coupling form factors defining the transitions of the positive and negative parity spin-3/2 pentaquark states. The values of the coupling constants are used in the calculation of the decay widths of these transitions. The obtained results  are compared with the existing experimental data.

\end{abstract}

\maketitle

The long and controversial history of pentaquark states has reached to a new stage with the announcement of the observation of two $P_c(4380)$ and $P_c(4450)$ states in 2015 by the LHCb Collaboration~\cite{Aaij:2015tga}. The possible quark substructure of these hadrons are different from the conventional hadrons composed of a quark and an antiquark or three quarks/antiquarks according to quark model. However, neither the quark model nor the QCD exclude the existence of these non-conventional hadrons. As a result of that, they have been investigated both theoretically and experimentally for very long time  to obtain indications for their existence. Finally these indications were attained by LHCb~\cite{Aaij:2015tga} putting them at the focus of interest. 

Considering the possibility of their existence, these types of states were studied extensively even before their observation in 2015. Their properties were investigated theoretically (see for instance Refs.~\cite{Jaffe:1976ig,Gignoux:1987cn,Hogaasen:1978jw,Strottman:1979qu,Lipkin:1987sk,
Fleck:1989ff,Oh:1994np,Chow,Shmatikov,Genovese,Lipkin2,Lichtenberg}). In the experimental side  for  a pentaquark state with quark content $uudd\bar{s}$ ($ \Theta^+ $), the observation was firstly claimed in 2003 in the interaction $\gamma n\rightarrow n K^+K^-$~\cite{Nakano:2003qx}. This claim was followed by the other experimental investigations \cite{Barmin:2003vv,Stepanyan:2003qr,Aktas:2004qf,Bai:2004gk,Knopfle:2004tu,Pinkenburg:2004ux,Harris:2004kx,Karshon:2004tf,Abt:2004tz,Aubert:2004bm,Litvintsev:2004yw,Karshon:2004kt,Link:2005ti,Aubert:2007qea} which ended up with either positive or negative signals leaving us with an ambiguity in their observation story. In the meantime there was an intense struggle in the theoretical side of researches to provide an explanation to experimental indications or provide insights into them [see the Ref.~\cite{Azizi:2016dhy} and the references therein].

In 2013 with the observation of $Z_c$~\cite{Ablikim:2013mio}, being an indication for the existence of pentaquark, the attentions have centered upon pentaquarks again. While there still were some null results coming from experimental researches such as the results of ALICE Collaboration investigating $\phi(1869)$ pentaquark~\cite{Abelev:2014qqa} and J-PARC E19 Collaboration searching for $\Theta^+$~\cite{Moritsu::2014qra}, the theoretical studies were indicating necessity for searching the pentaquark states with heavy quark contents \cite{Gerasyuta:2014wka}. Finally there came the long sought result  from the LHCb Collaboration with the announcement of  the observation of $P_c^+(4380)$ and $P_c^+(4450)$, in the
$\Lambda_b^0\rightarrow J/\psi K^-p$ decay with masses $4380\pm 8\pm 29$ MeV and $4449.8\pm 1.7\pm 2.5$ MeV, spins $3/2$ and $5/2$ and decay widths $205\pm 18\pm
86$ MeV and $39\pm 5\pm 19$ MeV, respectively~\cite{Aaij:2015tga}. That was followed by interpretation of other states as possible pentaquark states \cite{Kim:2017jpx,Yang:2017rpg,Huang:2017dwn,He:2017aps} such as some of the newly observed $\Omega_c$ states by LHCb~\cite{Aaij:2017nav}  and  the states $N(1875)$ and $N(2100)$.

After the observation of LHCb there have been intense theoretical works to explain the properties of these states. They were investigated through different models. The Ref.~\cite{Chen:2016qju} provides a review on these models which covers the multiquark states including pentaquarks and their  possible experimental measurements. They were investigated through meson-baryon molecular model~\cite{Yang:2015bmv,Burns:2015dwa,Lu:2016nnt,Tazimi:2016hsv,Wang:2016dzu,Shen:2016tzq,Roca:2015dva,Chen:2015loa,Huang:2015uda,Meissner:2015mza,Xiao:2015fia,
He:2015cea,Chen:2015moa,Wang:2015qlf,Chen:2016heh,Yamaguchi:2016ote,He:2016pfa}, diquark-triquark model~\cite{Wang:2016dzu,Zhu:2015bba,Lebed:2015tna}, diquark-diquark-antiquark
model~\cite{Wang:2016dzu,Anisovich:2015cia,Maiani:2015vwa,Ghosh:2015ksa,Wang:2015ava,Wang:2015epa,Wang:2015ixb},  and
topological soliton model~\cite{Scoccola:2015nia} to gain informations on the substructure and properties of them. Their properties were also studied using a variant of D4-D8 model~\cite{Liu:2017xzo}. The Refs.~\cite{yeni3,yeni4,yeni5,yeni6} discussed their being kinematical effects due to  triangle singularities. 

As previously mentioned, following the announcement of the observation of pentaquarks there have been extensive amount of works on their properties. However to gain a deep understanding on their nature and substructure, which are still not certain yet, we are in need of more experimental and theoretical investigations which may shed light on their properties. Studying their possible decay channels may provide valuable insights in this respect. One can find a few works in the literature which address their decay mechanisms. In Ref.~\cite{Wang:2015qlf} strong decay behaviors of these states were investigated considering them as  molecular states using the spin rearrangement scheme with heavy quark limit. With molecular state assumption for $P_c(4380)$,  one can find a rough estimation on the partial decay width of it in Ref.~\cite{Shen:2016tzq}. The strong decay mode $ J/\psi N $ was also studied in Refs.~\cite{Lu:2016nnt,Lin:2017mtz,Ortega:2016syt} which also take the structure of $ P_c(4380) $ and $ P_c(4450) $ in the molecular form.  Magnetic moment of the hidden-charm pentaquark state was studied in Ref.~\cite{Wang:2016dzu} for three models which are molecular model, diquark-diquark-antiquark and diquark-triquark models. This work resulted in different  magnetic moments for different configurations which indicate that experimental measurements on magnetic moments of them could be helpful for the determination of  their inner structure. In Ref.~\cite{Eides:2017xnt} the pentaquarks were interpreted as  hadroquarkonium states and the partial decay width of the  $P_c(4450)$ state, which could be explained by this model satisfactorily, was estimated as
$\Gamma( P_c(4450)\rightarrow N+J/\psi)\approx11$~MeV. Comparing the obtained result with the experimental data, the $P_c(4450)$ state was interpreted as a member of one of the two almost degenerate hidden-charm baryon octets with spin-parities $J^P=1/2^-, 3/2^-$.

As it can be seen from the references given above, different models could give results consistent with experimental results for the pentaquark masses. Therefore more works are needed to identify their inner structure. Investigation of their decay channels may be helpful in this respect. For that purpose, the investigation of the possible decay channels $\bar{D}\Sigma_c^*$, $J/\psi N$, $\bar{D}^*\Lambda_c$, $\bar{D}\Lambda_c$, $\eta_c N$, $ \eta_c \Delta $ and $ J/\psi \Delta $, may provide valuable insights. Considering this motivation in this work we study the decay channel $P_c(4380)\longrightarrow J/ \psi N$ for both the positive and negative parity states associated to $P_c(4380)$. To calculate the strong coupling constant we apply three point QCD sum rules~\cite{Shifman:1978bx,Shifman:1978by}. To this end, we use the results of our previous calculation~\cite{Azizi:2016dhy} for the masses and residues of these particles. These parameters are among the main input parameters in the calculation of the coupling constant of the considered decay. In Ref.~\cite{Azizi:2016dhy} the masses and residues were obtained for both positive and negative parity states of the considered pentaquarks. To fulfill the calculations the interpolating current is chosen in the molecular form $[\bar{D}^*\Sigma_{c}]$ for the $J=3/2$ pentaquark state.

To calculate the physical parameters in QCD sum rules, there are three steps that one follows. First one is the calculation of a correlator which
gives the phenomenological description of the correlator. The second one is the theoretical calculation of the same correlator using the operator
product expansion (OPE). And final stage is the match of the both descriptions which results in the physical parameters of the hadrons under consideration. To fulfill these steps the starting point is the construction of the suitable interpolating currents of the hadrons of interest. The present work deals with $P_c(4380)\longrightarrow J/ \psi N$ decay and these interpolating currents are
 \begin{eqnarray}
J_{\mu}^{P_{c}}&=&[\bar{c}_{d}\gamma_{\mu}d_{d}][\epsilon_{abc}(u_{a}^{T}C\gamma_{\theta}u_{b})\gamma^{\theta}\gamma_{5}c_{c}],
\nonumber\\
J^{N}&=&\epsilon_{abc} u_{a}^{T} C\gamma_{\mu}u_{b}\gamma_{5}\gamma^{\mu}d_{c},
\nonumber\\
J^{J/\psi}_{\mu}&=&\bar{c}\gamma_{\mu}c.
 \label{InterpolatingCurrents}
\end{eqnarray}
The three point correlation function that is used has the form
\begin{equation}
\Pi _{\mu \nu }(p, q)=i^2\int d^{4}xe^{-ip\cdot
x}\int d^{4}ye^{ip'\cdot y}\langle 0|\mathcal{T} \{J^{N}(y)
J_{\mu}^{J/\psi}(0)\bar{J}_{\nu}^{P_{c}}(x)\}|0\rangle,
\label{eq:CorrF1Pc}
\end{equation}
where $\mathcal{T}$ represents the time ordering product, $ J_{\nu}^{P_{c}} $, $ J^{N} $ and $ J^{J/\psi}_{\mu} $ are the interpolating currents given in Eq.~\ref{InterpolatingCurrents} which carry the same quantum numbers with the considered hadrons.
Here  we would like to make a remark on the current $ J_{\nu}^{P_{c}}$. Though the current $ J_{\nu}^{P_{c}}$ has a definite negative  parity, it interacts not only with the negative parity state but also with the positive parity one. This is due to the fact that the multiplication of the present current by $i\gamma_5$, that is $ i\gamma_5 J_{\nu}^{P_{c}}$, results in a reversion of the parity (see for instance the Refs.~\cite{Chung,Bagan,Jido,Wang:2015ava,Wang:2015epa,Wang:2015ixb} on this subject). Usage of a current in the form  $ i\gamma_5 J_{\nu}^{P_{c}}$ will not give us any independent sum rules from those that are obtained using the current $J_{\nu}^{P_{c}}$. In other words, the above correlation function contains  the information of both parity states.

In the physical side of the calculations the interpolating currents are treated as annihilation and creation operators of hadrons. Therefore this side leads us to the results in terms of hadronic degrees of freedom such as the masses and the coupling constants of the hadrons. For the calculation of this side we insert  complete sets of hadronic states having same quantum numbers with the interpolating currents into the correlation function and use the following definitions:
\begin{eqnarray}
\langle 0|J_{\nu }^{P_c}|P_c^{+}(p)\rangle &=&\lambda_{P_c^{+}} \gamma_5 u_{\nu}^{P_c^+}(p),\nonumber\\
\langle 0|J_{\nu }^{P_c}|P_c^{-}(p)\rangle &=&\lambda_{P_c^{-}} u_{\nu}^{P_c^-}(p).
\nonumber\\
\langle 0|J^{N }|N(p')\rangle &=&\lambda_{N} u^{N}(p'),
\nonumber\\
\langle 0|J_{\mu }^{J/\Psi}|J/\Psi(q)\rangle &=&f_{J/\Psi} m_{J/\Psi} \varepsilon_{\mu},
\label{eq:ResPc}
\end{eqnarray}
where $ f_{J/\Psi} $ is the decay constant and $ \varepsilon_{\mu} $ is the polarization vector of the $ J/\Psi $ state; and we use the superscripts  $ +$ and $-$ in the symbolization of $|P_c^{+(-)}\rangle  $ to represent the positive and negative parity states of spin-3/2 pentaquarks. Here, $\lambda_{P_c^{+(-)}} $ and $ \lambda_{N} $ are the residues of the $ P_c^{+(-)} $ and $ N $  states; and $ u_{\nu}^{P_c} $ and $ u^{N} $ are their spinors, respectively. After performing the four-integrals in Eq. (\ref{eq:CorrF1Pc}) we get
\begin{eqnarray}
\Pi _{\mu \nu }^{\mathrm{Phys}}(p, q)&=&\frac{\langle 0|J^{N
}|N(p')\rangle \langle 0|J_{\mu }^{J/\Psi}|J/\Psi(q)\rangle
\langle J/\Psi(q) N(p')|P_c^+(p)\rangle \langle P_c^+(p)|J_{\nu
}^{P_c}|0\rangle }{(m_N^2-p'^2)(m_{J/\Psi}^2-q^2)(m_{P_c^+}^2-p^2)}
\nonumber\\
&+&\frac{\langle 0|J^{N }|N(p')\rangle \langle 0|J_{\mu }^{J/\Psi}|J/\Psi(q)\rangle \langle J/\Psi(q) N(p')|P_c^-(p)\rangle \langle P_c^-(p)|J_{\nu }^{P_c}|0\rangle }{(m_N^2-p'^2)(m_{J/\Psi}^2-q^2)(m_{P_c^-}^2-p^2)}+\cdots,  \label{eq:CorrF1Phys}
\end{eqnarray}
where we included both the positive and negative parity ground state contributions as both couple to the same current. The $\cdots$ in Eq.~\ref{eq:CorrF1Phys}
represents the contributions coming from higher states and
continuum. In addition to the matrix elements defined in Eq. (\ref{eq:ResPc}), the following matrix elements defined in terms of the coupling constants $ g_i $  and $ \tilde{g}_i $ are also needed \cite{Aliev:2011kn}:
\begin{eqnarray}
\langle J/\Psi(q) N(p')|P_c^+(p)\rangle  &=&\bar{u}^{N}(p',s') \{g_1(q_{\alpha}
{\varepsilon^{*}}\!\!\!/-\varepsilon^{*}_{\alpha}q\!\!\!/)\gamma_5+g_2(P\cdot\varepsilon^{*} q_{\alpha}-P\cdot q \varepsilon^{*}_{\alpha})\gamma_5+g_3(q\cdot\varepsilon^{*} q_{\alpha}-q^2 \varepsilon^{*}_{\alpha})\gamma_5\}u_{\alpha}^{P_c^+}(p,s),
\nonumber\\
\langle J/\Psi(q) N(p')|P_c^-(p)\rangle  &=&\bar{u}^{N}(p',s') \{\tilde{g}_1(q_{\alpha}{\varepsilon^{*}}\!\!\!/-\varepsilon^{*}_{\alpha}{q}\!\!\!/)+\tilde{g}_2(P\cdot\varepsilon^{*} q_{\alpha}-P\cdot q \varepsilon^{*}_{\alpha})+\tilde{g}_3(q\cdot\varepsilon^{*} q_{\alpha}-q^2 \varepsilon^{*}_{\alpha})\}u_{\alpha}^{P_c^-}(p,s),
\label{eq:ResPc1}
\end{eqnarray}
with $P=\frac{p+p'}{2}$ and $ q=p-p' $.  With the substitution of the matrix elements into the
Eq.~\ref{eq:CorrF1Phys}, applying  summations over the polarization vector of $ J/\Psi $ meson as well as over the Dirac and
Rarita-Schwinger spinors using
\begin{eqnarray}
\sum_{s}u_{\mu}^{P_c^\pm}(p,s)\bar{u}_{\nu}^{P_c^\pm}(p,s)&=&-({\slashed
p}+m_{P_c^\pm})\left[ g_{\mu\nu}-\frac{1}{3}\gamma_{\mu} \gamma_{\nu}-
\frac{2p_{\mu}p_{\nu}}{3m_{P_c^\pm}^2}
+\frac{p_{\mu}\gamma_{\nu}-p_{\nu}\gamma_{\mu}}{3m_{P_c^\pm}}\right],\nonumber \\
\sum_{s'}u^{N}(p',s')\bar{u}^{N}(p',s')&=&({\slashed
p'}+m_{N}), \nonumber\\
\varepsilon_{\alpha}\varepsilon^*_{\beta}&=&-g_{\alpha\beta}+\frac{q_\alpha q_\beta}{m_{J/\Psi}^2},
\label{eq:SumPc}
\end{eqnarray}
and performing Borel transformation with the aim of suppressing the contributions of the higher states and continuum, we obtain the final form of the correlation function in the
physical side as:
\begin{eqnarray}
\mathcal{B}\Pi_{\mu \nu
}^{\mathrm{Phys}}(q)&=&e^{-\frac{m_N^2}{M^{'2}}}\frac{f_{J/\psi}\lambda_{N}
m_{J/\psi}} {q^2-m_{J/\psi}^2}\left\{\left[g_1
\Phi_1+\tilde{g}_1\tilde{\Phi}_1 \right]\gamma_{\mu}p'_{\nu}
+\left[g_1 \Phi_2+\tilde{g}_1\tilde{\Phi}_2
\right]p\!\!\!/\gamma_{\mu} p'_{\nu}+ \left[
g_1\frac{2\Phi_2}{3m_N} +\tilde{g}_1
\frac{2\tilde{\Phi}_2}{3m_N}\right.\right.
 \nonumber \\
 &-&\left. g_2\frac{\Phi_1-3m_N m_{P_c^+}}{12 m_{P_c^+}}+\tilde{g}_2
 \frac{\tilde{\Phi}_1
 +3m_Nm_{P_c^-}}{12m_{P_c^-}}+g_3\frac{\Phi_1-3m_N m_{P_c^+}}{6 m_{P_c^+}}
 -\tilde{g}_3\frac{\tilde{\Phi}_1+3m_Nm_{P_c^-}
 }{6m_{P_c^-}}
\right]p\!\!\!/ p'\!\!\!\!/\gamma_{\nu}p'_{\mu}
 \nonumber \\
 &-& \left[
 g_2\frac{\Phi_1-2m_Nm_{P_c^+}}{2}+
 \tilde{g}_2\frac{\tilde{\Phi}_1+2m_Nm_{P_c^-}}{2}
 -g_3(\Phi_1-2m_Nm_{P_c^+})
-
 \tilde{g}_3(\tilde{\Phi}_1+2m_Nm_{P_c^-})
 \right]p'_{\mu}p'_{\nu}
\nonumber \\
 &-&\left[g_1\frac{\Phi_2(m_N+m_{P_c^+})}{3m_{P_c^+}}-\tilde{g}_1
 \frac{\tilde{\Phi}_2
 (m_N-m_{P_c^-})}
 {3m_{P_c^-}}+g_2\frac{m_N(\Phi_1+m_Nm_{P_c^+}-4m_{P_c^+}^2+2q^2)}{12m_{P_c^+}}
 \right.
\nonumber \\
&-&
\tilde{g}_2\frac{m_N(\tilde{\Phi}_1-m_Nm_{P_c^-}-4m_{P_c^-}^2+2q^2)}{12m_{P_c^-}}
-g_3\frac{m_N(\Phi_1-3m_Nm_{P_c^+}+2q^2)}{6m_{P_c^+}}
\nonumber \\
&+&\left.\left.
\tilde{g}_3\frac{m_N(\Phi_1+3m_Nm_{P_c^-}+2q^2)}{6m_{P_c^-}}
 \right]p\!\!\!/\gamma_{\nu}p_{\mu}
+
 \left[g_2 \frac{\Phi_2}{2m_N}+ \tilde{g}_2 \frac{\tilde{\Phi}_2}{2m_N}
 +g_3 \frac{\Phi_2}{m_N}+ \tilde{g}_3 \frac{\tilde{\Phi}_2}{m_N}\right]
 p\!\!\!/ p'\!\!\!\!/p_{\mu}p'_{\nu}
\right\}
\nonumber \\
&+&\mathrm{other \,\,\, structures}
 +\textellipsis\ ,
\label{eq:CorBorPc}
\end{eqnarray}
where
\begin{eqnarray}
\Phi_1&=& \lambda_{P_c^+}
\left(m_N^2+m_Nm_{P_c^+}+m_{P_c^+}^2-q^2\right)
e^{-\frac{m_{P_c^+}^2}{M^{2}}},\nonumber \\
\tilde{\Phi}_1&=& \lambda_{P_c^-}
\left(m_N^2-m_Nm_{P_c^-}+m_{P_c^-}^2-q^2\right)
e^{-\frac{m_{P_c^-}^2}{M^{2}}},
\nonumber \\
\Phi_2&=& \lambda_{P_c^+} m_N e^{-\frac{m_{P_c^+}^2}{M^{2}}},
\nonumber \\
\tilde{\Phi}_2&=& \lambda_{P_c^-} m_N
e^{-\frac{m_{P_c^-}^2}{M^{2}}},
 \label{eq:Coeff}
\end{eqnarray}
and $ M^2 $ and $ M'^2 $ are Borel parameters to be fixed later.

The calculation in the theoretical side is done inserting explicit expressions of the interpolating currents into the correlation function. Contraction of the quark fields via Wick's theorem leads us to the result
\begin{eqnarray}
\Pi_{\mu \nu }^{OPE}(p,p',q)&=&i^2\int d^{4}xe^{-ip\cdot x}\int d^{4}ye^{ip'\cdot y}\epsilon^{abc}\epsilon^{a'b'c'}\bigg\{-Tr[\gamma^{\theta}CS_{u}^{Tba'}(y-x)C\gamma^{\beta}S_{u}^{ab'}(y-x)]\gamma_5 \gamma_{\beta}S_d^{cd'}(y-x) \gamma_{\nu} \nonumber\\&\times&S_{c}^{d'd}(x) \gamma_{\mu} S_{c}^{dc'}(-x)\gamma_5 \gamma_{\theta}+Tr[\gamma^{\theta}CS_{u}^{Taa'}(y-x)C\gamma^{\beta}S_{u}^{bb'}(y-x)]\gamma_5 \gamma_{\beta}S_d^{cd'}(y-x) \gamma_{\nu} \nonumber\\&\times&S_{c}^{d'd}(x) \gamma_{\mu} S_{c}^{dc'}(-x)\gamma_5 \gamma_{\theta}\bigg\},\label{eq:CorrF1Theore}
\end{eqnarray}
where $a,b,c,...$ are color indices, $C$ is charge conjugation operator; and $S_{u,d}$ and $S_{c}$ are the light and heavy quark propagators whose explicit expressions are as follows
\begin{eqnarray}
 S_{q}^{ab}(x)=&&i\frac{x\!\!\!/}{2\pi^{2}x^{4}}\delta_{ab}-\frac{m_{q}}{4\pi^{2}x^{2}}\delta_{ab}-\frac{\langle
 \overline{q}q\rangle}{12}\Big(1-i\frac{m_{q}}{4}x\!\!\!/\Big)\delta_{ab}-\frac{x^{2}}{192}m_{0}^{2}\langle
 \overline{q}q\rangle\Big( 1-i\frac{m_{q}}{6}x\!\!\!/\Big)\delta_{ab}\\
 &&
 \notag-\frac{ig_{s}G_{ab}^{\theta\eta}}{32\pi^{2}x^{2}}\Big[x\!\!\!/\sigma_{\theta\eta}+\sigma_{\theta\eta}x\!\!\!/ \Big]-\frac{x\!\!\!/ x^{2}g_s^2}{7776}\langle
 \overline{q}q\rangle^2\delta_{ab}+\cdots,
\end{eqnarray}
\begin{eqnarray}
\notag
 S_{c}^{ab}(x)=&&\frac{i}{(2\pi)^{4}}\int
 d^{4}ke^{-ik.x}\Big\{\frac{\delta_{ab}}{k\!\!\!/-m_{c}}-\frac{g_{s}G_{ab}^{\alpha\beta}}{4}\frac{\sigma_{\alpha\beta}(k\!\!\!/+m_{c})+(k\!\!\!/+m_{c})\sigma_{\alpha\beta}}{(k^{2}-m_{c}^{2})^{2}}\\
 &&+\frac{\pi^{2}}{3}\Big\langle\frac{\alpha_{s}GG}{\pi}\Big\rangle\delta_{ab}m_{c}\frac{k^{2}+m_{c}k\!\!\!/}{(k^{2}-m_{c}^{2})^{4}}+\cdots\Big\}.
\end{eqnarray}

The correlation function $\Pi_{\mu \nu }^{OPE}(p,q)$ has again terms containing different Dirac structures and can be written as
\begin{eqnarray}
\Pi_{\mu \nu }^{OPE}(p,q)&=&\Pi_1\, \gamma_{\mu}p'_{\nu} +
\Pi_2\, p\!\!\!/\gamma_{\mu} p'_{\nu} + \Pi_3\, p\!\!\!/
p'\!\!\!\!/\gamma_{\nu}p'_{\mu}+ \Pi_4\, p'_{\mu}p'_{\nu} +
\Pi_5\, p\!\!\!/\gamma_{\nu}p_{\mu} + \Pi_6\, p\!\!\!/
p'\!\!\!\!/p_{\mu}p'_{\nu}\nonumber \\
&+&\mathrm{other \,\,\, structures}. \label{eq:PiOPE}
\end{eqnarray}
The $ \Pi_i $ functions given in the Eq.~\ref{eq:PiOPE} are calculated  via substitution of quark propagators into Eq.~\ref{eq:CorrF1Theore}. This is followed by the transformation of the calculations to the momentum space. The imaginary parts of the results give us the spectral densities which are used in the following
dispersion integral to obtain the final results of the OPE side
\begin{eqnarray}
\Pi_{i}=\int ds\int
ds'\frac{\rho_{i}^{pert}(s,s',q^{2})+\rho_{i}^{non-pert}(s,s',q^{2})}{(s-p^{2})
(s'-p'^{2})}. \label{eq:Pispect}
\end{eqnarray}
where $i=1,2,..,6$; and $ \rho_{i}^{pert}(s,s',q^{2}) $ and $ \rho_{i}^{non-pert}(s,s',q^{2}) $ are the perturbative and non-perturbative parts of the spectral densities, respectively. All these steps summarized above result in lengthy expressions for the spectral densities. In order not to overwhelm the study with overlong mathematical expressions   we prefer not to present them here.

After the calculations of hadronic and OPE sides, to get the QCD sum rules, we choose and match the coefficients of the same structures from both sides to obtain the coupling constants entering the calculations. We have six coupling constants and here we only present two of them,         $g_1$ and $\tilde{g}_1$, to provide insight into the forms of the others which have more or less similar forms,
\begin{eqnarray}
g_1&=&
e^{\frac{m_{P_c^+}^2}{M^2}}e^{\frac{m_N^2}{M'^2}}\frac{(m_{J/\psi}^2-q^2)
\left[\Pi_2(m_N^2-m_Nm_{P_c^-}+m_{P_c^-}^2-q^2) -m_N\Pi_1 \right]}
{f_{J/\psi}\lambda_N\lambda_{P_c^+}m_Nm_{J/\psi}(m_{P_c^+}+m_{P_c^-})
(m_N+m_{P_c^+}-m_{P_c^-})}\nonumber \\
\tilde{g}_1&=&-
e^{\frac{m_{P_c^-}^2}{M^2}}e^{\frac{m_N^2}{M'^2}}\frac{(m_{J/\psi}^2-q^2)
\left[\Pi_2(m_N^2+m_Nm_{P_c^+}+m_{P_c^+}^2-q^2) -m_N\Pi_1 \right]}
{f_{J/\psi}\lambda_N\lambda_{P_c^-}m_Nm_{J/\psi}(m_{P_c^+}+m_{P_c^-})
(m_N+m_{P_c^+}-m_{P_c^-})}.
  \label{eq:g1}
\end{eqnarray}

\begin{table}[tbp]
\begin{tabular}{|c|c|}
\hline\hline Parameters & Values \\ \hline\hline
$m_{c}$ & $(1.67\pm0.07)~\mathrm{GeV}$ \cite{PDG}\\
$m_{P_c^+}$ & $(4.24\pm0.16)~\mathrm{GeV}$ \cite{Azizi:2016dhy}\\
$m_{P_c^-}$ & $(4.30\pm0.10)~\mathrm{GeV}$ \cite{Azizi:2016dhy}\\
$m_{J/\psi}$ & $(3096.900\pm0.006)~\mathrm{MeV}$ \cite{PDG}\\
$m_{N}$ & $(938.272081\pm0.000006)~\mathrm{MeV}$ \cite{PDG}\\
$\lambda_{P_c^+}$ & $(0.59\pm0.07)\times 10^{-3}~\mathrm{GeV}^6$ \cite{Azizi:2016dhy}\\
$\lambda_{P_c^-}$ & $(0.94\pm0.05)\times 10^{-3}~\mathrm{GeV}^6$ \cite{Azizi:2016dhy}\\
$\lambda_{N}^2$ & $(0.0011\pm0.0005)~\mathrm{GeV}^6$ \cite{Azizi:2014yea}\\
$f_{J/\psi}$ & $(481\pm36)~\mathrm{MeV}$ \cite{Veliev:2011kq}\\
$\langle \bar{q}q \rangle $ & $(-0.24\pm 0.01)^3$ $\mathrm{GeV}^3$  \\
$m_{0}^2 $ & $(0.8\pm0.1)$ $\mathrm{GeV}^2$ \\
$\langle \overline{q}g_s\sigma Gq\rangle$ & $m_{0}^2\langle
\bar{q}q \rangle$
\\
$\langle\frac{\alpha_sG^2}{\pi}\rangle $ & $(0.012\pm0.004)$ $~\mathrm{GeV}%
^4 $\\
\hline\hline
\end{tabular}%
\caption{Some input parameters used in the calculations.}
\label{tab:Param}
\end{table}

The  sum rules for the strong coupling constants involve some input parameters that are required to obtain the behaviors of the coupling constants as a function of $Q^2 = -q^2$. These input parameters are presented in Table~\ref{tab:Param}. In the calculations the light quark masses, $ m_u $ and $ m_d $ are set to zero. The above input parameters are not the only ones required. There exist four more auxiliary parameters: the continuum thresholds $s_{0}$,  $s'_{0}$, appearing after applying the continuum subtractions according to the standard prescriptions, and Borel parameters $M^{2}$ and $ M'^{2} $. We need to establish them before going further. To specify their working intervals we require weak dependencies of the physical quantities that we aim to obtain on these parameters and   follow some necessary criteria. For determination of the intervals of Borel parameters we consider the adequate suppression of  higher states and continuum and demand the convergence of the OPE as being a series expansion. Our analysis on results considering these requirements eventuate in the intervals as follows:
\begin{eqnarray}
5.5\ \mathrm{GeV}^{2}\leq M^{2}& \leq & 7.5\ \mathrm{GeV}^{2},
\nonumber \\
1.0\ \mathrm{GeV}^{2}\leq M'^{2}& \leq & 1.5\ \mathrm{GeV}^{2}.
\label{Eq:MsqMpsq}
\end{eqnarray}%

The continuum thresholds are not completely arbitrary, but they depends on the energies of the first excited states in the initial and final channels.  In the nucleon channel, the first excited state is well established experimentally. We choose the value of $ s'_0 $ such that the first excited state in nucleon channel has not been included in the calculations. However, in $ P_c $ channel, we have unfortunately no sufficient experimental information on the energy of the first excited state, hence, we choose the region of the continuum threshold $ s_0 $ in accordance with the information on the spectrum of the standard charmed baryons and such that the dependence on it be relatively weak. Besides these points, the OPE convergence and maximum possible pole contribution are our main demands that should be satisfied. These considerations lead to the intervals,
\begin{eqnarray}
22.0\,\,\mathrm{GeV}^{2}&\leq& s_{0}\leq24.0\,\,\mathrm{GeV}^{2},
\nonumber \\
1.8\,\,\mathrm{GeV}^{2}&\leq& s'_{0}\leq2.1\,\,\mathrm{GeV}^{2}.
\label{Eq:s0s0p}
\end{eqnarray}
\begin{figure}[h!]
\begin{center}
\includegraphics[totalheight=5cm,width=7cm]{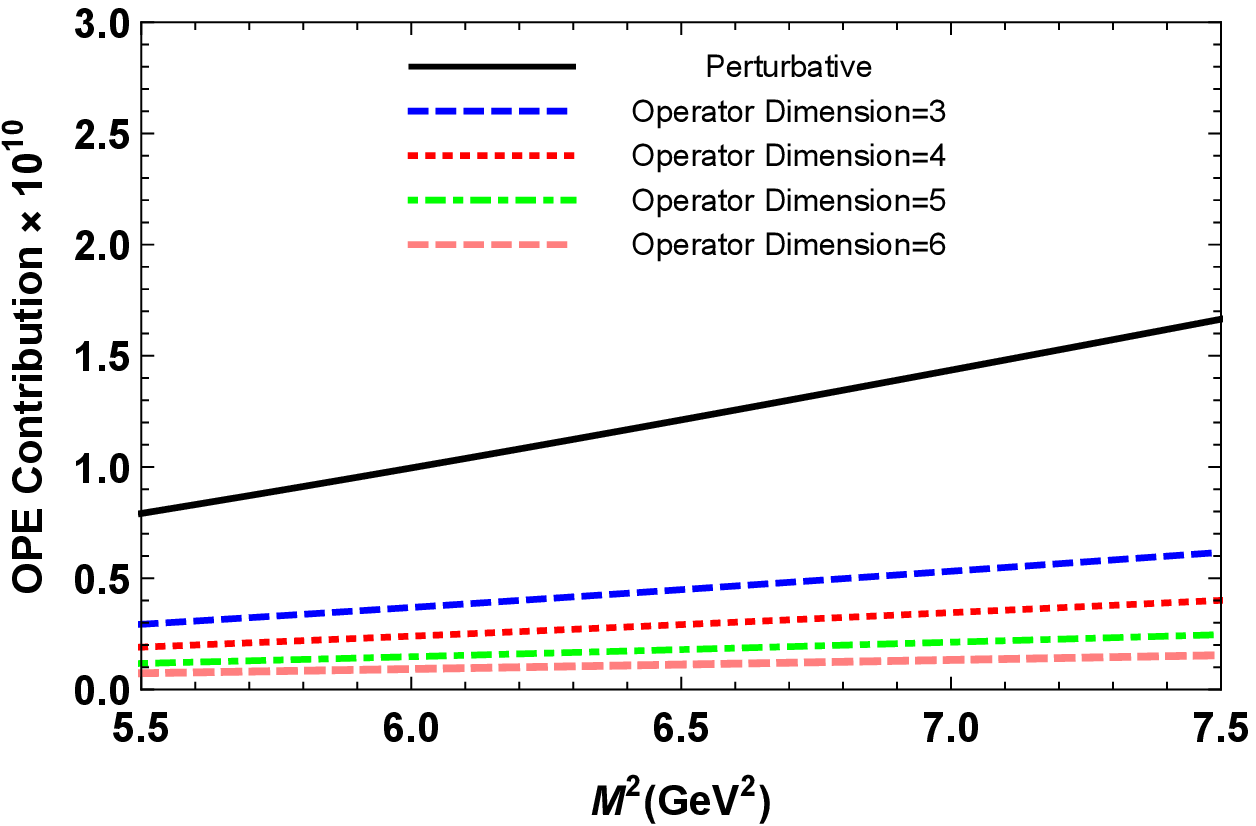}
\includegraphics[totalheight=5cm,width=7cm]{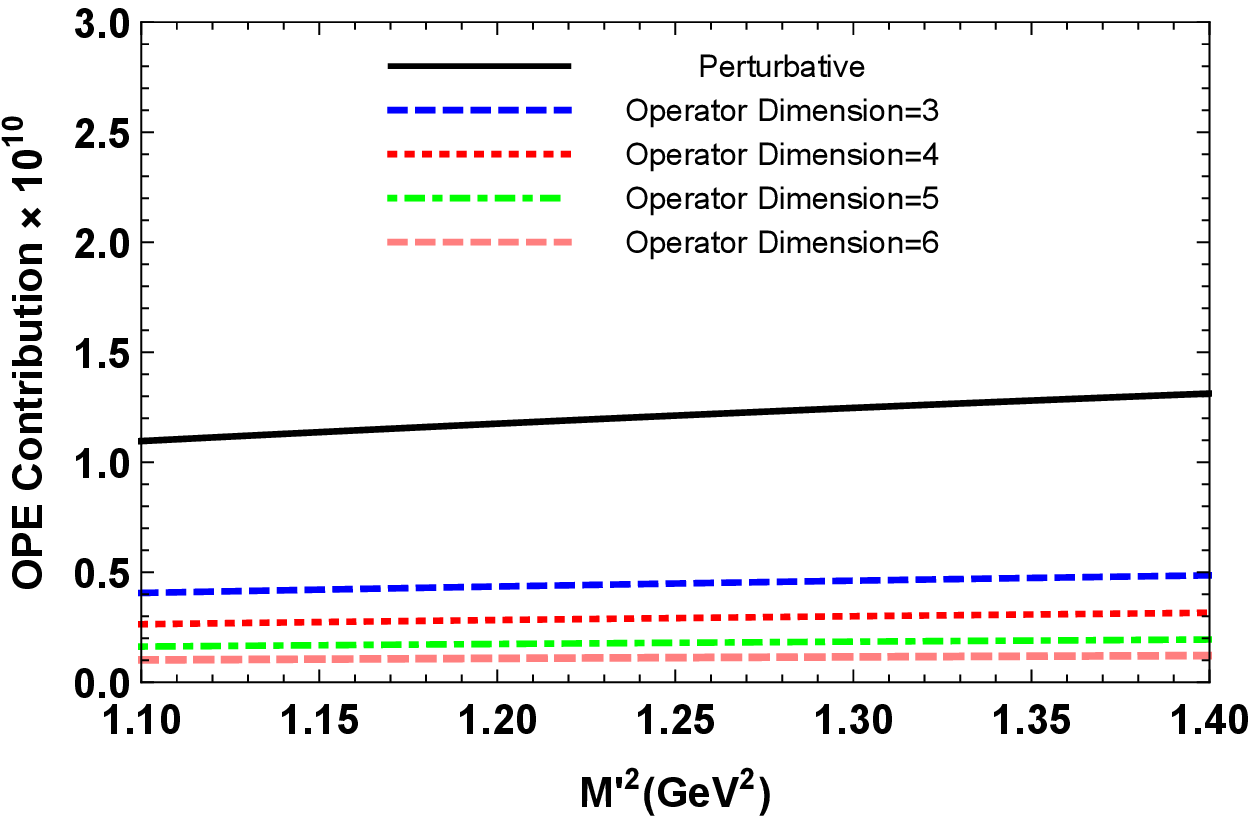}
\end{center}
\caption{\textbf{Left:} Contributions of the perturbative and different nonperturbative operators to OPE as a function of $M^2$ with central
values of other auxiliary parameters and  at $Q^2=4~\mathrm{GeV^2}$.
\textbf{Right:} The same as left panel but as a function of  $M'^2$. }
\label{gr:OPEvsMsqMpsq}
\end{figure}

\begin{figure}[h!]
\begin{center}
\includegraphics[totalheight=5cm,width=7cm]{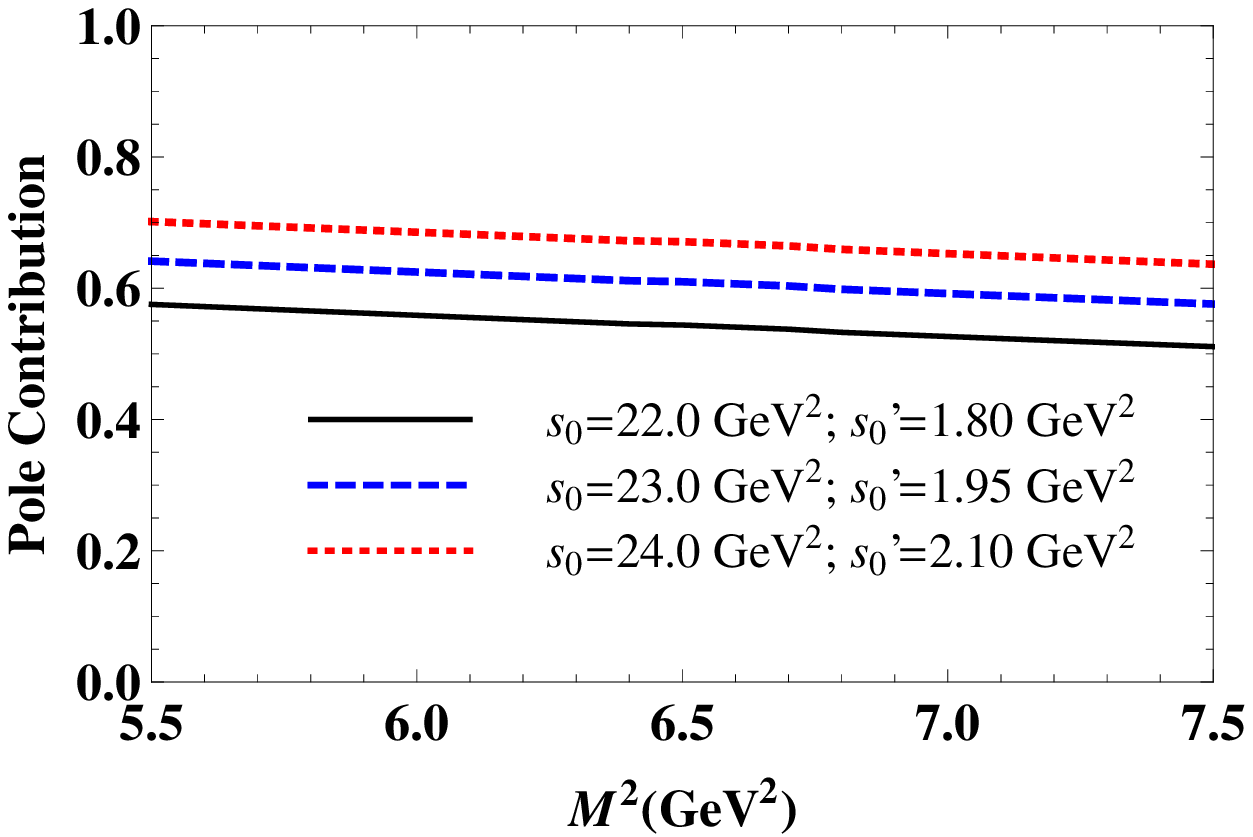}
\includegraphics[totalheight=5cm,width=7cm]{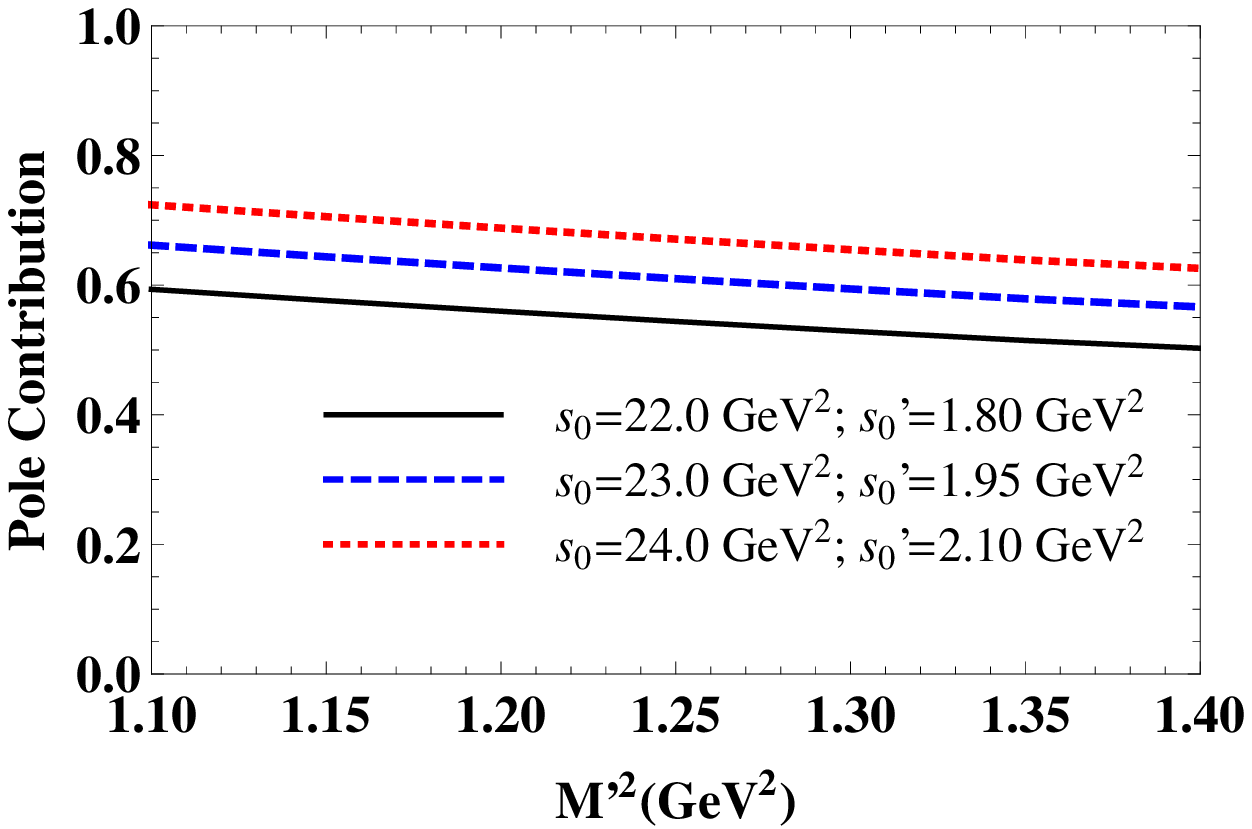}
\end{center}
\caption{\textbf{Left:} The pole contribution as a function of $M^2$ at various threshold parameters $s_0$ and $s_0'$ .
\textbf{Right:} The same as left panel, but  as a function of $M'^2$.  }
\label{gr:PCvsMsqMpsq}
\end{figure}
To show how the conditions of the OPE convergence and pole dominance are satisfied we plot their dependence on the auxiliary parameters, as an example for the structure $ p'_{\mu}p'_{\nu} $, in their working regions in Figs.~\ref{gr:OPEvsMsqMpsq} and \ref{gr:PCvsMsqMpsq}. From these figures, we see that the standard prescriptions and transformations to sufficiently suppress the contributions of the higher states and continuum and enhance the pole contributions have ended up in a very good OPE convergence and a good pole dominance. Such that the contribution of the perturbative part exceeds the total nonperturbative contributions and when the dimension of operators is increased the contribution decreases leading to a nice OPE convergence. From figure \ref{gr:PCvsMsqMpsq} it is clear that the pole contribution constitutes more than  $ 50\% $ of the total contributions in the working windows of auxiliary parameters. 

\begin{figure}[h!]
\begin{center}
\includegraphics[totalheight=5cm,width=7cm]{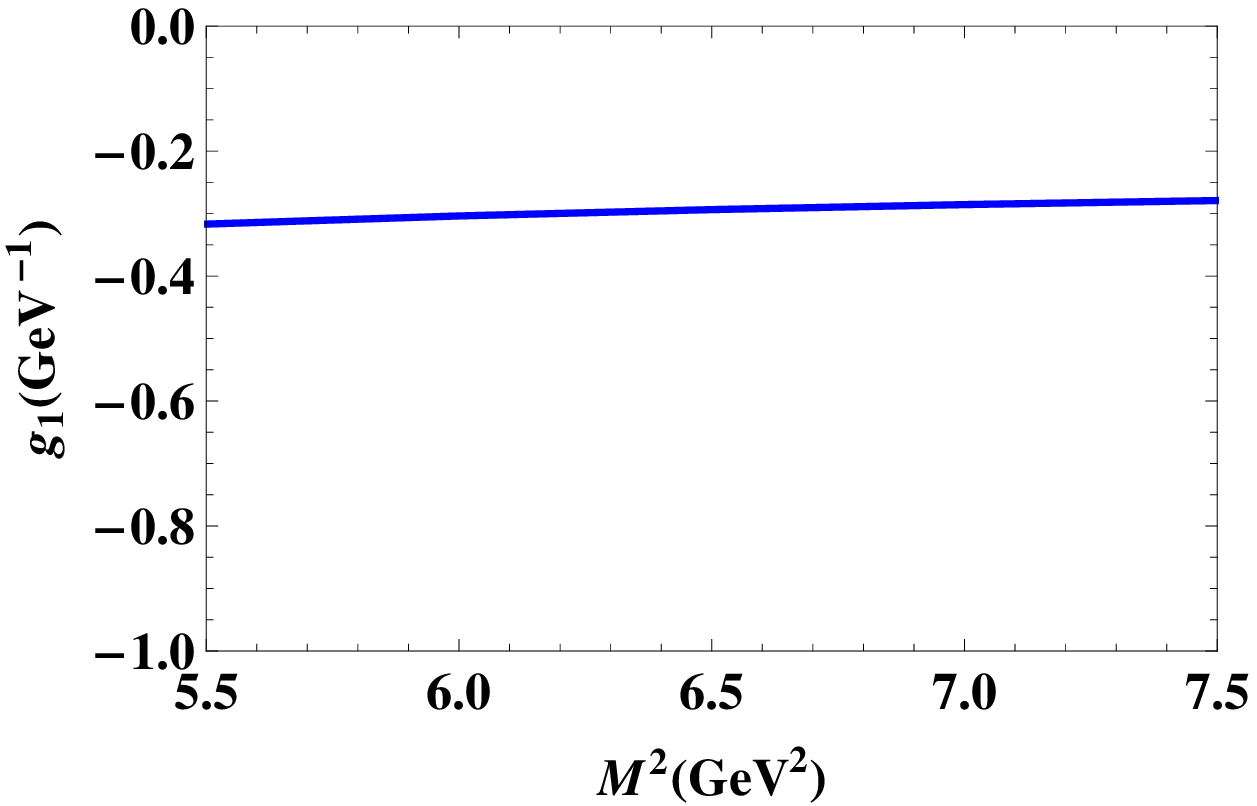}
\includegraphics[totalheight=5cm,width=7cm]{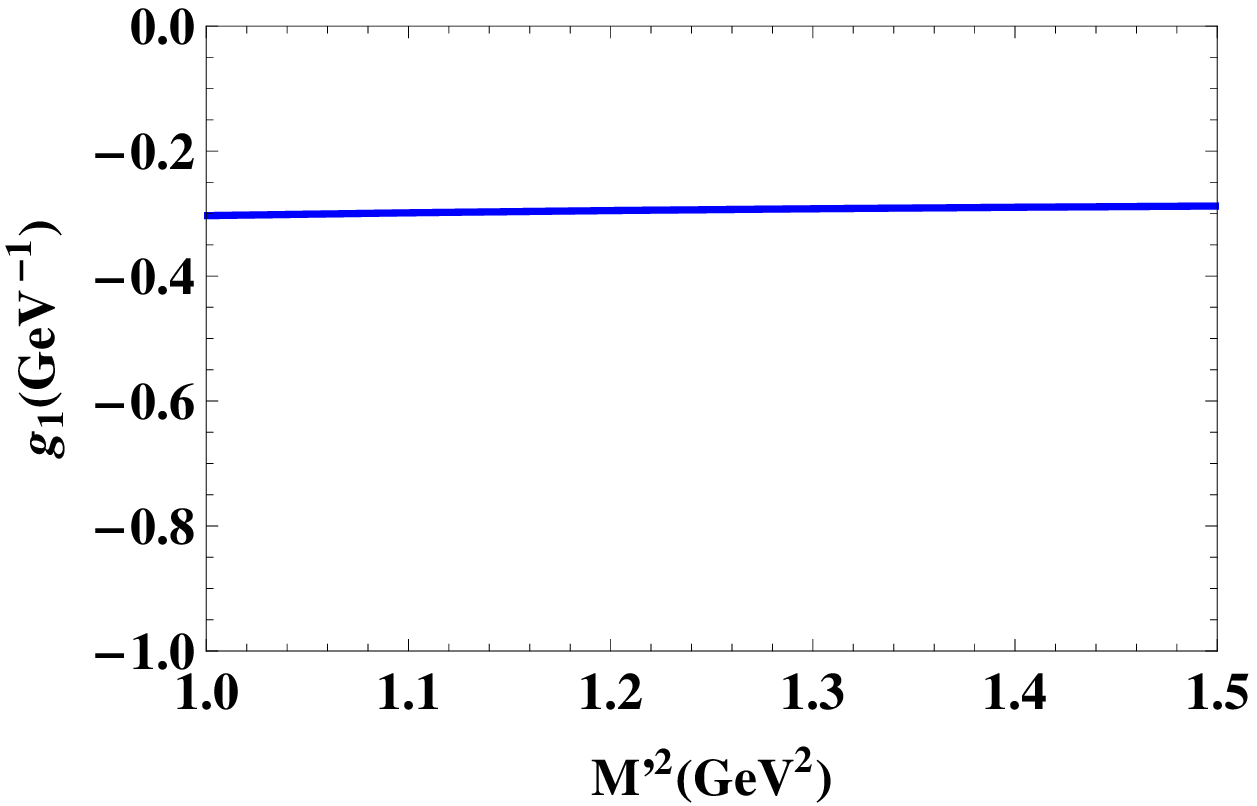}
\end{center}
\caption{\textbf{Left:} $g_1$ as a function of $M^2$ with central
values of other auxiliary parameters and $Q^2=4~\mathrm{GeV^2}$.
\textbf{Right:} The same as left panel but as a function of $M'^2$. }
\label{gr:g1vsMsqMpsq}
\end{figure}

\begin{figure}[h!]
\begin{center}
\includegraphics[totalheight=5cm,width=7cm]{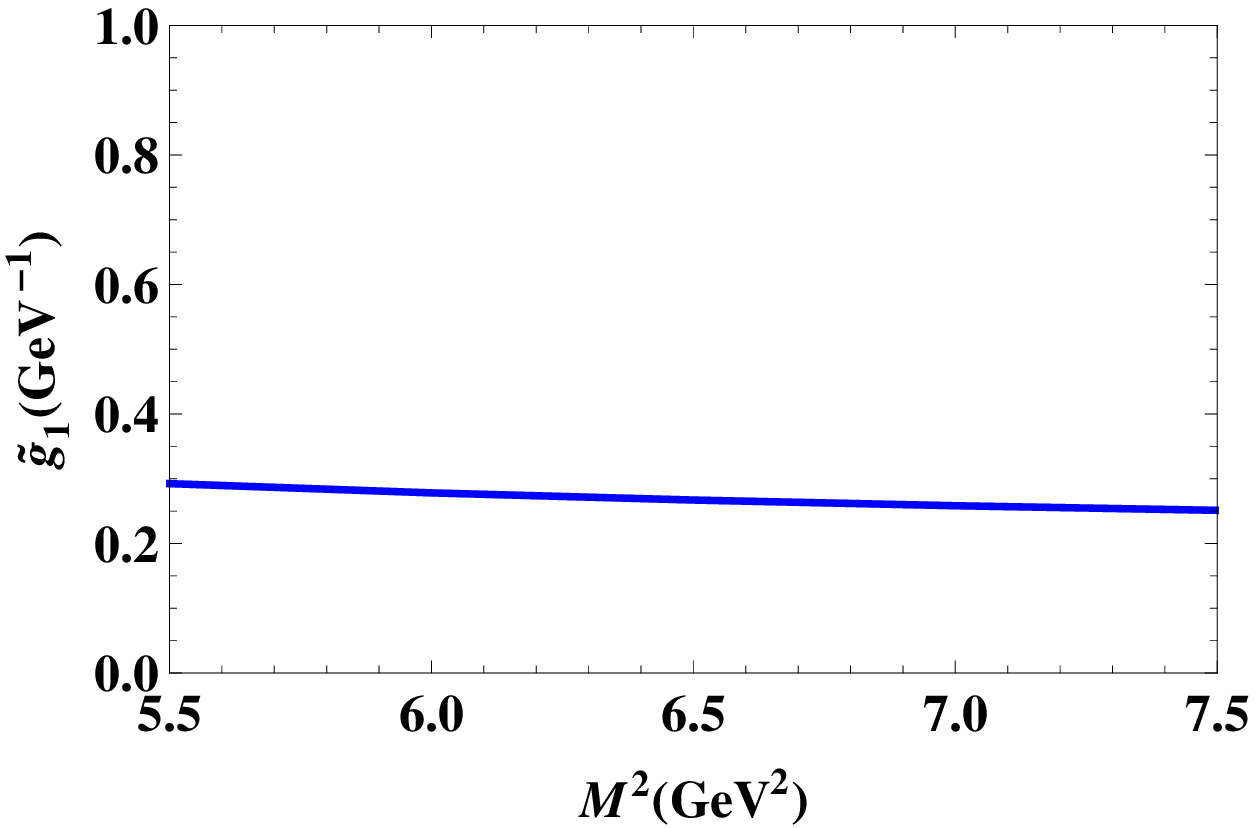}
\includegraphics[totalheight=5cm,width=7cm]{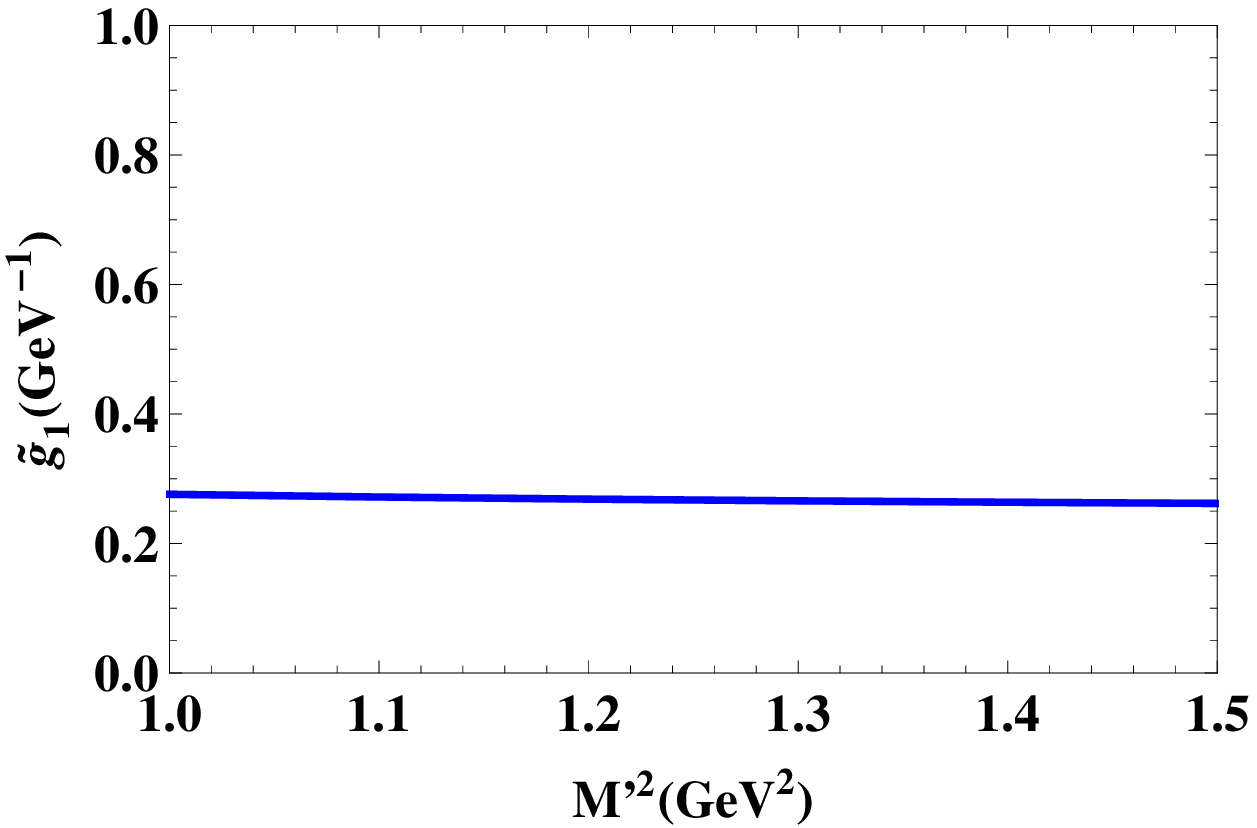}
\end{center}
\caption{\textbf{Left:}  The same as Fig. 1 but $\tilde{g}_1$ as a function of $M^2$.
\textbf{Right:} The same as Fig. 1 but  $\tilde{g}_1$ as a function of $M'^2$.  }
\label{gr:g1pvsMsqMpsq}
\end{figure}

Using the working intervals of the auxiliary parameters we plot the dependencies of $g_1$ and $\tilde{g}_1$ on the  Borel parameters, $M^2$ and $M'^2$,  in Figs.~\ref{gr:g1vsMsqMpsq} and \ref{gr:g1pvsMsqMpsq},
respectively. In these figures, for the threshold parameters we apply their central values. From these figures one can see that, as required, there are a mild dependencies of the results on the auxiliary parameters. In theory one expects the results be completely independent of these parameters but in practice, though small, there appear some dependency. As a consequence, this weak dependency of the results on the threshold parameters and Borel masses  brings the main sources of uncertainty to the computation.   

\begin{figure}[h!]
\begin{center}
\includegraphics[totalheight=5cm,width=7cm]{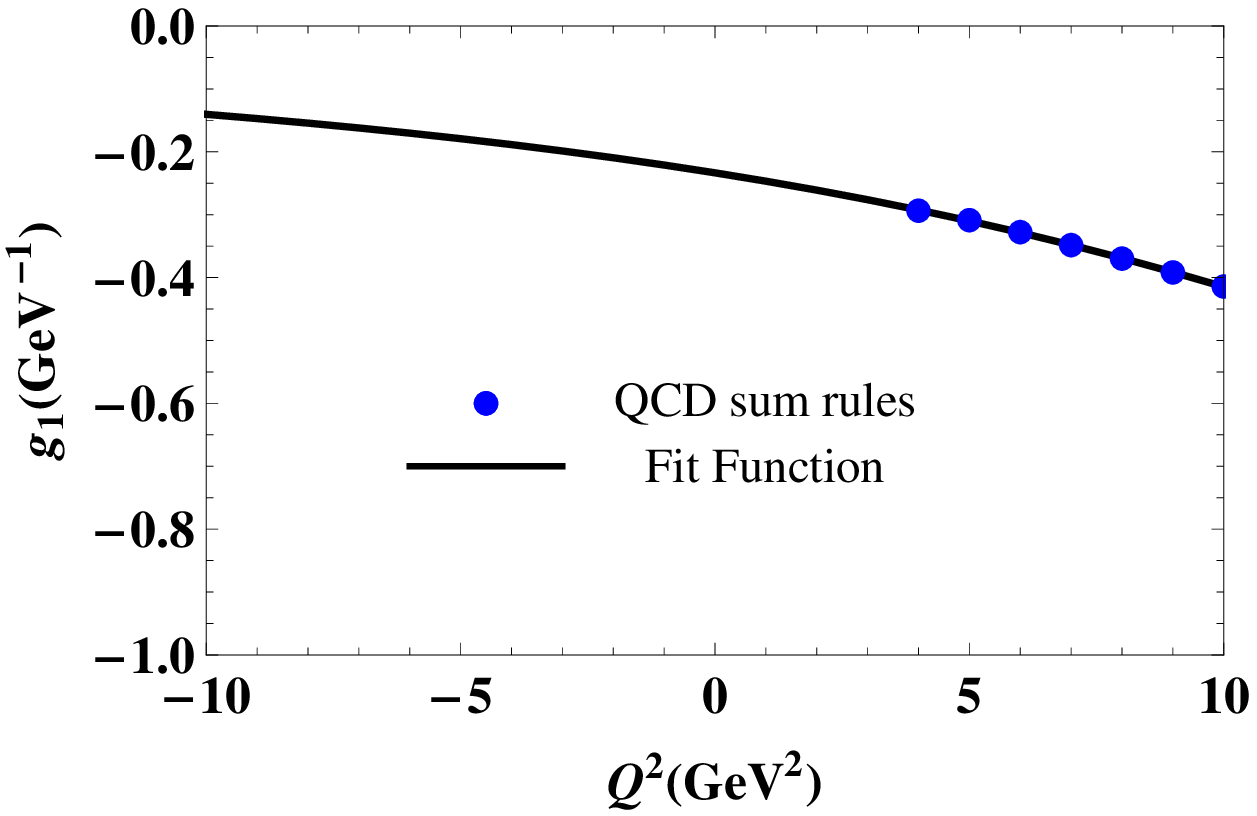}
\includegraphics[totalheight=5cm,width=7cm]{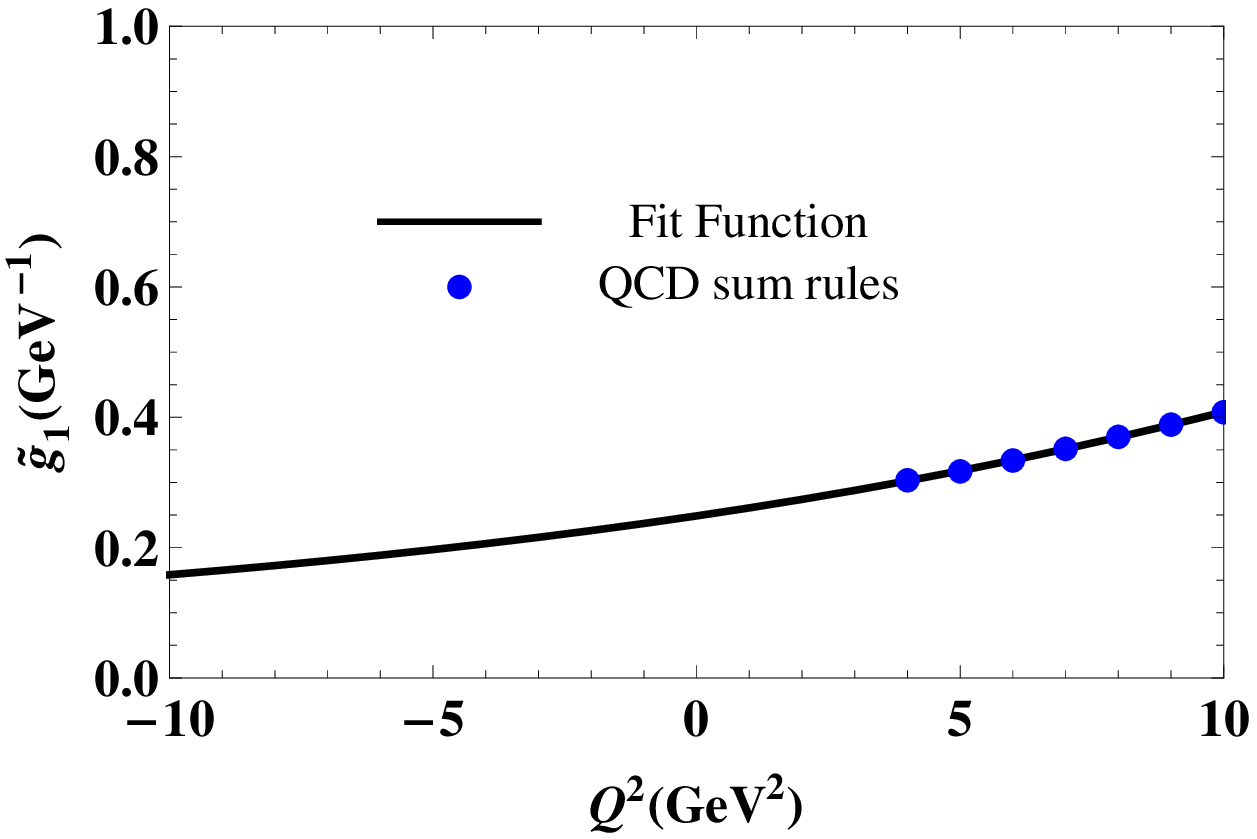}
\end{center}
\caption{\textbf{Left:} $g_1$ as a function of $Q^2$.
\textbf{Right:} $\tilde{g}_1$ as a function of $Q^2$.  }
\label{gr}
\end{figure}

After the determination of working intervals of the auxiliary parameters, we apply them together with  other input parameters to obtain the dependence of the coupling form factors on    $Q^2$. To represent the results the following fit functions
are applied
\begin{eqnarray}
g_i(Q^2)&=&\frac{g_0}{1-a(\frac{Q^2}{m_{P_c^+}^2})+b(\frac{Q^2}{m_{P_c^+}^2})^2},
\nonumber \\
\tilde{g_i}(Q^2)&=&\frac{g_0}{1-a(\frac{Q^2}{m_{P_c^-}^2})+b(\frac{Q^2}{m_{P_c^-}^2})^2}.
\end{eqnarray}
where $g_0$, $a$ and $b$ are the fit parameters having the values presented in Table~\ref{tab:FitParam} for  coupling form factors under consideration. To exemplify the consistency of the fit functions with our sum rule results we present Figure~\ref{gr} which shows the
dependencies of the strong coupling constants, $ g_1 $ and $\tilde{g}_1$ , on
$Q^2$ obtained from both sum rules and fit results. This figure indicates that the chosen fit functions represent QCD sum rule results well in the region where our sum rule results are reliable. Therefore to obtain the coupling constants at $Q^2 = -m_{J/\psi}^2$ we use the fit functions and obtain the values of strong coupling constants for considered
transitions as presented in Table~\ref{tab:CouplingValues}. The errors in the results are due to the errors coming from the determination of the working regions of the auxiliary parameters as well as the uncertainties of other input parameters.
\begin{table}[tbp]
\begin{tabular}{|c|c|c|c|}
\hline\hline Coupling Constant  &$ g_0$ & a & b \\ \hline\hline
 $g_1$&$-0.232~(\mathrm{GeV}^{-1})$& 0.989 & 0.336 \\
  $\tilde{g}_1$&$0.249~(\mathrm{GeV}^{-1})$ & 0.891 & 0.329 \\
   $g_2$&$2.876~(\mathrm{GeV}^{-2})$ & 2.452 & 2.716 \\
     $\tilde{g}_2$&$-0.944~(\mathrm{GeV}^{-2})$& 3.338 & 4.655 \\
    $g_3$&$1.007~(\mathrm{GeV}^{-2} )$& 2.427 & 2.001 \\
     $\tilde{g}_3$&$-0.576~(\mathrm{GeV}^{-2})$ &3.374 & 4.589 \\
\hline\hline
\end{tabular}%
\caption{: Parameters appearing in the fit functions of the
coupling constants.} \label{tab:FitParam}
\end{table}

\begin{table}[tbp]
\begin{tabular}{|c|c|c|c|c|c|}
\hline\hline $g_1~(\mathrm{GeV}^{-1})$  &$
\tilde{g}_1~(\mathrm{GeV}^{-1})$ & $g_2~(\mathrm{GeV}^{-2})$
  &$ \tilde{g}_2~(\mathrm{GeV}^{-2})$&$g_3~(\mathrm{GeV}^{-2})$  &
  $ \tilde{g}_3~(\mathrm{GeV}^{-2})$ \\
\hline\hline
$-0.15\pm0.03$&$0.17\pm0.04$&$0.93\pm0.25$&$-0.24\pm0.07$&$0.35\pm0.08$&$-0.15\pm0.04$\\
\hline\hline
\end{tabular}
\caption{: Values of the strong coupling constants.}
\label{tab:CouplingValues}
\end{table}

As a final task in this work we calculate corresponding widths for the decay channels $P_c^+ \longrightarrow J/\psi N$ and $P_c^-
\longrightarrow J/\psi N$ following the standard methods
and using the definitions for the strong couplings  defined in Eq.~\ref{eq:ResPc} as well as parameters of the involved particles. Our calculations result in:
\begin{eqnarray}
\Gamma(P_c^+ \longrightarrow
J/\psi N)&=&\frac{f(m_{P_c^+},m_{J/\psi},m_N)}{384\pi
m_{P_c^+}^4}\left[(m_{P_c^+}-m_N)^2-m_{J/\psi}^2\right]
 \nonumber \\
&\times&
 \left[g_1^2\phi_1+g_2^2\phi_2
+g_3^2\phi_3+g_1\,g_2\,\phi_4+g_1\,g_3\,\phi_5+g_2\,g_3\,\phi_6\right],
\label{Eq:DWPozPart}
\end{eqnarray}
for positive parity and 
\begin{eqnarray}
\Gamma(P_c^- \longrightarrow
J/\psi N)&=&\frac{f(m_{P_c^-},m_{J/\psi},m_N)}{384\pi
m_{P_c^-}^4}\left[(m_{P_c^-}+m_N)^2-m_{J/\psi}^2\right]
 \nonumber \\
&\times&
 \left[\tilde{g}_1^2\tilde{\phi}_1+\tilde{g}_2^2\tilde{\phi}_2
+\tilde{g}_3^2\tilde{\phi}_3+\tilde{g}_1\,\tilde{g}_2\,\tilde{\phi}_4+\tilde{g}_1\,
\tilde{g}_3\,\tilde{\phi}_5+\tilde{g}_2\,\tilde{g}_3\,\tilde{\phi}_6\right],
\label{Eq:DWNegativeParity}
\end{eqnarray}
for negative parity, where
\begin{eqnarray}
f(x,y,z)&=&\frac{1}{2x}\sqrt{x^4+y^4+z^4-2xy-2xy-2yz},
 \nonumber \\
\phi_1&=&8\left[m_{J/\psi}^4-2m_{J/\psi}^2\left(m_N^2-m_{P_c^+}^2+m_{P_c^+}m_N
+(m_{P_c^+}+m_N)^2(3m_{P_c^+}^2+m_N^2)\right)\right],
\nonumber \\
\phi_2&=&m_{J/\psi}^6+m_{P_c^+}^2\left(3m_{P_c^+}^2+m_N^2\right)\left(3m_{P_c^+}^2
+m_N^2
-2m_{J/\psi}^2\right)+8m_{P_c^+}^2\left(m_{P_c^+}^2-m_N^2\right)^2,
\nonumber \\
\phi_3&=&4m_{J/\psi}^2\left[m_{J/\psi}^2\left(m_{J/\psi}^2+10m_{P_c^+}^2-2m_N^2\right)+
\left(m_{P_c^+}^2-m_N^2\right)^2\right],
\nonumber \\
\phi_4&=&-8m_{P_c^+}\left[m_{J/\psi}^2\left(m_{J/\psi}^2-4m_{P_c^+}^2\right)+
\left(m_{P_c^+}+m_N\right)^2\left(3m_{P_c^+}^2+m_N^2-4m_{P_c^+}m_N\right)\right],
\nonumber \\
\phi_5&=&32m_{P_c^+}m_{J/\psi}^2\left(m_{J/\psi}^2+2m_{P_c^+}^2-m_N^2+m_{P_c^+}m_N
\right),
\nonumber \\
\phi_6&=&
-4m_{J/\psi}^2\left[m_{J/\psi}^4+m_N^4-11m_{P_c^+}^4+10m_{P_c^+}^2m_N^2-2m_{J/\psi}^2
\left(m_{P_c^+}^2+m_N^2\right)\right],
\nonumber \\
\tilde{\phi}_1&=&8\left[m_{J/\psi}^4+2m_{J/\psi}^2\left(-m_N^2+m_{P_c^-}^2+m_{P_c^-}m_N
+(m_{P_c^-}-m_N)^2(3m_{P_c^-}^2+m_N^2)\right)\right],
\nonumber \\
\tilde{\phi}_2&=&m_{J/\psi}^6+m_{P_c^-}^2\left(3m_{P_c^-}^2+m_N^2\right)
\left(3m_{P_c^-}^2+m_N^2
-2m_{J/\psi}^2\right)+8m_{P_c^-}^2\left(m_{P_c^-}^2-m_N^2\right)^2,
\nonumber \\
\tilde{\phi}_3&=&4m_{J/\psi}^2\left[m_{J/\psi}^2\left(m_{J/\psi}^2+10m_{P_c^-}^2-2m_N^2
\right)+ \left(m_{P_c^-}^2-m_N^2\right)^2\right],
\nonumber \\
\tilde{\phi}_4&=&-8m_{P_c^-}\left[m_{J/\psi}^2\left(m_{J/\psi}^2-4m_{P_c^-}^2\right)-
\left(m_{P_c^-}-m_N\right)^2\left(3m_{P_c^-}^2+m_N^2+4m_{P_c^-}m_N\right)\right],
\nonumber \\
\tilde{\phi}_5&=&32m_{P_c^-}m_{J/\psi}^2\left(m_{J/\psi}^2+2m_{P_c^-}^2-m_N^2+m_{P_c^-}
m_N \right),
\nonumber \\
\tilde{\phi}_6&=&
-4m_{J/\psi}^2\left[m_{J/\psi}^4+m_N^4-11m_{P_c^-}^4+10m_{P_c^-}^2m_N^2-2m_{J/\psi}^2
\left(m_{P_c^-}^2+m_N^2\right)\right].
 \label{Eq:DWCoef}
\end{eqnarray}
Using these  formulas and the obtained values for the coupling constants, we find
\begin{eqnarray}
\Gamma(P_c^+ \longrightarrow J/\psi N)&=&\left(187.3\pm53.8\right)
\mathrm{MeV},
\nonumber \\
\Gamma(P_c^- \longrightarrow J/\psi N)&=&\left(212.5\pm60.4\right)
\mathrm{MeV}.
 \label{Eq:DWNegativeParity}
\end{eqnarray}
As is seen, the obtained  central value of the decay width for the negative parity case is relatively close to the reported central   experimental value for the width of the $ P_c (4380) $ state previously presented in the text. However, when we consider the theoretical as well as the experimental uncertainties our predictions for the widths of both parities overlap with the experimental data. 

To sum up, in this work we considered the $ P_c \longrightarrow J/\psi N $ decay channels of both the  positive and negative parity pentaquark states. For the considered transitions the strong coupling constants were calculated via three point QCD sum rule method using a current having molecular form of $[\bar{D}^*\Sigma_{c}]$ . The results attained for six coupling constants were used for the calculation of the decay widths of the corresponding channels. In the literature one can find different assumptions for the inner structures of the observed pentaquark states. Based on these assumptions the masses of these hadrons are obtained to be nicely consistent with the experimental observations. Therefore to distinguish the different models and find the best suggestion for the substructure of these states, it is necessary to provide further theoretical studies on other physical parameters of these states. Among these studies are their strong decay mechanisms. In this respect this study may provide valuable insights into their substructure and shed light on future experiments. 
Combination of the results of the present study on the widths of the positive- and negative-parity $ P_c $ 
decaying to $ J/\psi N $ with our predictions on their masses using the same  picture from Ref. \cite{Azizi:2016dhy}, may help experimentalists to establish the internal quark organization of the observed $ P_c (4380) $ and uniquely determine its parity.    One may consider the strong decays of the positive- and negative-parity $ P_c(4450) $ states to $ J/\psi N $ to compare with the available experimental data. For this, the corresponding matrix elements in terms of the strong coupling form factors should be defined.

\section*{ACKNOWLEDGEMENTS}

K. A. and Y. S. thank  T\"{U}B\.{I}TAK for partial support provided under the Grant no: 115F183.

\label{sec:Num}


\begin{thebibliography}{99}

\bibitem{Aaij:2015tga}
  R.~Aaij {\it et al.} [LHCb Collaboration],
  Phys.\ Rev.\ Lett.\  {\bf 115}, 072001 (2015).


\bibitem{Jaffe:1976ig}
  R.~L.~Jaffe,
  Phys.\ Rev.\ D {\bf 15}, 267; 281 (1977).


\bibitem{Gignoux:1987cn}
  C.~Gignoux, B.~Silvestre-Brac and J.~M.~Richard,
  Phys.\ Lett.\ B {\bf 193}, 323 (1987).

\bibitem{Hogaasen:1978jw}
  H.~Hogaasen and P.~Sorba,
  Nucl.\ Phys.\ B {\bf 145}, 119 (1978).

\bibitem{Strottman:1979qu}
  D.~Strottman,
  Phys.\ Rev.\ D {\bf 20}, 748 (1979).

\bibitem{Lipkin:1987sk}
  H.~J.~Lipkin,
  Phys.\ Lett.\ B {\bf 195}, 484 (1987).

\bibitem{Fleck:1989ff}
  S.~Fleck, C.~Gignoux, J.~M.~Richard and B.~Silvestre-Brac,
  Phys.\ Lett.\ B {\bf 220}, 616 (1989).

\bibitem{Oh:1994np}
  Y.~S.~Oh, B.~Y.~Park and D.~P.~Min,
  Phys.\ Lett.\ B {\bf 331}, 362 (1994).

\bibitem{Chow}
  C.~K.~Chow,
  Phys.\ Rev.\ D {\bf 51}, 6327 (1995).

\bibitem{Shmatikov}
  M.~Shmatikov,
  Nucl.\ Phys.\ A {\bf 612}, 449 (1997).

\bibitem{Genovese}
  M.~Genovese, J.~M.~Richard, F.~Stancu and S.~Pepin,
  Phys.\ Lett.\ B {\bf 425}, 171 (1998).

\bibitem{Lipkin2}
  H.~J.~Lipkin,
  Nucl.\ Phys.\ A {\bf 625}, 207 (1997).

\bibitem{Lichtenberg}
D.~B.~Lichtenberg,
J. Phys. G {\bf 24}, 2065 (1998).



\bibitem{Nakano:2003qx}
  T.~Nakano {\it et al.} [LEPS Collaboration],
  Phys.\ Rev.\ Lett.\  {\bf 91}, 012002 (2003).




\bibitem{Barmin:2003vv}
  V.~V.~Barmin {\it et al.} [DIANA Collaboration],
  Phys.\ Atom.\ Nucl.\  {\bf 66}, 1715 (2003)
  [Yad.\ Fiz.\  {\bf 66}, 1763 (2003)].

  \bibitem{Stepanyan:2003qr}
  S.~Stepanyan {\it et al.} [CLAS Collaboration],
  Phys.\ Rev.\ Lett.\  {\bf 91}, 252001 (2003).

  \bibitem{Aktas:2004qf}
  A.~Aktas {\it et al.} [H1 Collaboration],
  Phys.\ Lett.\ B {\bf 588}, 17 (2004).

  \bibitem{Bai:2004gk}
  J.~Z.~Bai {\it et al.} [BES Collaboration],
  Phys.\ Rev.\ D {\bf 70}, 012004 (2004).

\bibitem{Knopfle:2004tu}
  K.~T.~Knopfle {\it et al.} [HERA-B Collaboration],
  J.\ Phys.\ G {\bf 30}, S1363 (2004).

\bibitem{Pinkenburg:2004ux}
  C.~Pinkenburg [PHENIX Collaboration],
  J.\ Phys.\ G {\bf 30}, no. 8, S1201 (2004)

\bibitem{Harris:2004kx}
  F.~A.~Harris [BES Collaboration],
  Int.\ J.\ Mod.\ Phys.\ A {\bf 20}, 445 (2005).

\bibitem{Karshon:2004tf}
  U.~Karshon [ZEUS Collaboration],
  hep-ex/0407004.

\bibitem{Abt:2004tz}
  I.~Abt {\it et al.} [HERA-B Collaboration],
  Phys.\ Rev.\ Lett.\  {\bf 93}, 212003 (2004).

\bibitem{Aubert:2004bm}
  B.~Aubert {\it et al.} [BaBar Collaboration],
  hep-ex/0408064.

\bibitem{Litvintsev:2004yw}
  D.~O.~Litvintsev [CDF Collaboration],
  Nucl.\ Phys.\ Proc.\ Suppl.\  {\bf 142}, 374 (2005).

\bibitem{Karshon:2004kt}
  U.~Karshon [ZEUS Collaboration],
  hep-ex/0410029.

\bibitem{Link:2005ti}
  J.~M.~Link {\it et al.} [FOCUS Collaboration],
  Phys.\ Lett.\ B {\bf 622}, 229 (2005).

  \bibitem{Aubert:2007qea}
  B.~Aubert {\it et al.} [BaBar Collaboration],
  Phys.\ Rev.\ D {\bf 76}, 092004 (2007).
  
  
  



\bibitem{Azizi:2016dhy}
  K.~Azizi, Y.~Sarac and H.~Sundu,
  Phys.\ Rev.\ D {\bf 95}, no. 9, 094016 (2017)
  [arXiv:1612.07479 [hep-ph]].



  \bibitem{Ablikim:2013mio}
  M.~Ablikim {\it et al.} [BESIII Collaboration],
  Phys.\ Rev.\ Lett.\  {\bf 110}, 252001 (2013).

  \bibitem{Abelev:2014qqa}
  B.~B.~Abelev {\it et al.} [ALICE Collaboration],
  Eur.\ Phys.\ J.\ C {\bf 75}, no. 1, 1 (2015).


  \bibitem{Moritsu::2014qra}
  M.~Moritsu {\it et al.} [J-PARC E19 Collaboration],
  Phys.\ Rev.\ C {\bf 90}, no. 3, 035205 (2014).



\bibitem{Gerasyuta:2014wka}
  S.~M.~Gerasyuta, V.~I.~Kochkin and X.~Liu,
  Phys.\ Rev.\ D {\bf 91}, no. 5, 054037 (2015).


  \bibitem{Kim:2017jpx}
  H.~C.~Kim, M.~V.~Polyakov and M.~Praszałowicz,
  arXiv:1704.04082 [hep-ph].

  \bibitem{Yang:2017rpg}
  G.~Yang and J.~Ping,
  arXiv:1703.08845 [hep-ph].

  \bibitem{Huang:2017dwn}
  H.~Huang, J.~Ping and F.~Wang,
  arXiv:1704.01421 [hep-ph].

  \bibitem{He:2017aps} 
  J.~He,
  Phys.\ Rev.\ D {\bf 95}, no. 7, 074031 (2017).


   \bibitem{Aaij:2017nav}
  R.~Aaij {\it et al.} [LHCb Collaboration],
  arXiv:1703.04639 [hep-ex].



\bibitem{Chen:2016qju}
  H.~X.~Chen, W.~Chen, X.~Liu and S.~L.~Zhu,
  Phys.\ Rept.\  {\bf 639}, 1 (2016).






 \bibitem{Yang:2015bmv} G.~Yang and J.~Ping,
  Phys.\ Rev.\ D {\bf 95}, no. 1, 014010 (2017).

  \bibitem{Burns:2015dwa}
  T.~J.~Burns,
  Eur.\ Phys.\ J.\ A {\bf 51}, no. 11, 152 (2015).


\bibitem{Lu:2016nnt}
  Q.~F.~L\"{u} and Y.~B.~Dong,
  Phys.\ Rev.\ D {\bf 93}, no. 7, 074020 (2016).


  \bibitem{Tazimi:2016hsv}
  M.~Monemzadeh, N.~Tazimiand and S.~Babaghodrat,
  Adv.\ High Energy Phys.\  {\bf 2016}, 6480926 (2016).



\bibitem{Wang:2016dzu}
  G.~J.~Wang, R.~Chen, L.~Ma, X.~Liu and S.~L.~Zhu,
  Phys.\ Rev.\ D {\bf 94}, no. 9, 094018 (2016).

\bibitem{Shen:2016tzq}
  C.~W.~Shen, F.~K.~Guo, J.~J.~Xie and B.~S.~Zou,
  Nucl.\ Phys.\ A {\bf 954}, 393 (2016).

  \bibitem{Roca:2015dva}
  L.~Roca, J.~Nieves and E.~Oset,
  Phys.\ Rev.\ D {\bf 92}, no. 9, 094003 (2015).

\bibitem{Chen:2015loa}
  R.~Chen, X.~Liu, X.~Q.~Li and S.~L.~Zhu,
  Phys.\ Rev.\ Lett.\  {\bf 115}, no. 13, 132002 (2015).

\bibitem{Huang:2015uda}
  H.~Huang, C.~Deng, J.~Ping and F.~Wang,
  Eur.\ Phys.\ J.\ C {\bf 76}, no. 11, 624 (2016).

\bibitem{Meissner:2015mza}
  U.~G.~Mei\ss ner and J.~A.~Oller,
  Phys.\ Lett.\ B {\bf 751}, 59 (2015).

  \bibitem{Xiao:2015fia}
  C.~W.~Xiao and U.-G.~Mei\ss ner,
  Phys.\ Rev.\ D {\bf 92}, no. 11, 114002 (2015).

\bibitem{He:2015cea}
  J.~He,
  Phys.\ Lett.\ B {\bf 753}, 547 (2016).

\bibitem{Chen:2015moa}
  H.~X.~Chen, W.~Chen, X.~Liu, T.~G.~Steele and S.~L.~Zhu,
  Phys.\ Rev.\ Lett.\  {\bf 115}, no. 17, 172001 (2015).



\bibitem{Wang:2015qlf}
  G.~J.~Wang, L.~Ma, X.~Liu and S.~L.~Zhu,
  Phys.\ Rev.\ D {\bf 93}, no. 3, 034031 (2016).




\bibitem{Chen:2016heh}
  R.~Chen, X.~Liu and S.~L.~Zhu,
  Nucl.\ Phys.\ A {\bf 954}, 406 (2016).

\bibitem{Yamaguchi:2016ote} 
  Y.~Yamaguchi and E.~Santopinto,
  Phys.\ Rev.\ D {\bf 96}, no. 1, 014018 (2017).

\bibitem{He:2016pfa}
  J.~He,
  Phys.\ Rev.\ D {\bf 95}, no. 7, 074004 (2017)


  \bibitem{Zhu:2015bba}
  R.~Zhu and C.~F.~Qiao,
  Phys.\ Lett.\ B {\bf 756}, 259 (2016).

  \bibitem{Lebed:2015tna}
  R.~F.~Lebed,
  Phys.\ Lett.\ B {\bf 749}, 454 (2015).


  \bibitem{Anisovich:2015cia}
  V.~V.~Anisovich, M.~A.~Matveev, J.~Nyiri, A.~V.~Sarantsev and A.~N.~Semenova,
  arXiv:1507.07652 [hep-ph].

\bibitem{Maiani:2015vwa}
  L.~Maiani, A.~D.~Polosa and V.~Riquer,
  Phys.\ Lett.\ B {\bf 749}, 289 (2015).

\bibitem{Ghosh:2015ksa}
  R.~Ghosh, A.~Bhattacharya and B.~Chakrabarti,
  arXiv:1508.00356 [hep-ph].

\bibitem{Wang:2015ava}
  Z.~G.~Wang and T.~Huang,
  Eur.\ Phys.\ J.\ C {\bf 76}, no. 1, 43 (2016).

\bibitem{Wang:2015epa}
  Z.~G.~Wang,
  Eur.\ Phys.\ J.\ C {\bf 76}, no. 2, 70 (2016).

\bibitem{Wang:2015ixb}
  Z.~G.~Wang,
  Nucl.\ Phys.\ B {\bf 913}, 163 (2016).

  \bibitem{Scoccola:2015nia}
  N.~N.~Scoccola, D.~O.~Riska and M.~Rho,
  Phys.\ Rev.\ D {\bf 92}, no. 5, 051501 (2015).


\bibitem{Liu:2017xzo} 
  Y.~Liu and I.~Zahed,
  Phys.\ Rev.\ D {\bf 95}, no. 11, 116012 (2017).





\bibitem{yeni3}
F.~K.~Guo, U.~G.~Meisner, W.~Wang and Z.~Yang,
  Phys.\ Rev.\ D {\bf 92}, no. 7, 071502 (2015).



\bibitem{yeni4}
X.~H.~Liu, Q.~Wang and Q.~Zhao,
  Phys.\ Lett.\ B {\bf 757}, 231 (2016).



\bibitem{yeni5}
F.~K.~Guo, U.~G.~Meisner, J.~Nieves and Z.~Yang,
  Eur.\ Phys.\ J.\ A {\bf 52}, no. 10, 318 (2016).



  \bibitem{yeni6}

M.~Bayar, F.~Aceti, F.~K.~Guo and E.~Oset,
  Phys.\ Rev.\ D {\bf 94}, no. 7, 074039 (2016).
  
  
  \bibitem{Lin:2017mtz} 
  Y.~H.~Lin, C.~W.~Shen, F.~K.~Guo and B.~S.~Zou,
  Phys.\ Rev.\ D {\bf 95}, no. 11, 114017 (2017).

\bibitem{Ortega:2016syt} 
  P.~G.~Ortega, D.~R.~Entem and F.~Fernández,
  Phys.\ Lett.\ B {\bf 764}, 207 (2017).




\bibitem{Eides:2017xnt}
  M.~I.~Eides, V.~Y.~Petrov and M.~V.~Polyakov,
  arXiv:1709.09523 [hep-ph].

\bibitem{Shifman:1978bx}
  M.~A.~Shifman, A.~I.~Vainshtein and V.~I.~Zakharov,
  Nucl.\ Phys.\ B {\bf 147}, 385 (1979).
\bibitem{Shifman:1978by}
  M.~A.~Shifman, A.~I.~Vainshtein and V.~I.~Zakharov,
  Nucl.\ Phys.\ B {\bf 147}, 448 (1979).
  
  
  
  
  

\bibitem{Chung}
Y. Chung, H. G. Dosch, M. Kremer and D. Schall, Nucl. Phys. B \textbf{197} (1982) 55.

\bibitem{Bagan}
E. Bagan, M. Chabab, H. G. Dosch and S. Narison, Phys. Lett. B \textbf{301}, 243 (1993).

\bibitem{Jido}
D. Jido, N. Kodama and M. Oka, Phys. Rev. D54 (1996) 4532.



\bibitem{Aliev:2011kn} 
  T.~M.~Aliev, K.~Azizi, M.~Savci and V.~S.~Zamiralov,
  Phys.\ Rev.\ D {\bf 83}, 096007 (2011).










\bibitem{PDG}
C. Patrignani et al. (Particle Data Group), Chin. Phys. C, {\bf
40}, 100001 (2016) and 2017 update.

\bibitem{Azizi:2014yea}
  K.~Azizi and N.~Er,
  Eur.\ Phys.\ J.\ C {\bf 74}, 2904 (2014)

\bibitem{Veliev:2011kq}
  E.~V.~Veliev, K.~Azizi, H.~Sundu, G.~Kaya and A.~Turkan,
  Eur.\ Phys.\ J.\ A {\bf 47}, 110 (2011)


\end{thebibliography}
\end{document}